\documentclass[final]{agujournal2019}
\usepackage{url} %this package should fix any errors with URLs in refs.
\usepackage{lineno}
\usepackage[inline]{trackchanges} %for better track changes. finalnew option will compile document with changes incorporated.
\usepackage{soul}
\usepackage{xcolor}
\usepackage{colortbl}

\usepackage{amsmath}

\linenumbers

\let\linenumbers\nolinenumbers\nolinenumbers
\draftfalse

\journalname{Journal of Advances in Modeling Earth Systems (JAMES)}

\begin{document}

%% ------------------------------------------------------------------------ %%
%  Title
%
% (A title should be specific, informative, and brief. Use
% abbreviations only if they are defined in the abstract. Titles that
% start with general keywords then specific terms are optimized in
% searches)
%
%% ------------------------------------------------------------------------ %%

% Example: \title{This is a test title}

\title{Adjoint-based online learning of two-layer quasi-geostrophic baroclinic turbulence}

%% ------------------------------------------------------------------------ %%
%
%  AUTHORS AND AFFILIATIONS
%
%% ------------------------------------------------------------------------ %%

% Authors are individuals who have significantly contributed to the
% research and preparation of the article. Group authors are allowed, if
% each author in the group is separately identified in an appendix.)

% List authors by first name or initial followed by last name and
% separated by commas. Use \affil{} to number affiliations, and
% \thanks{} for author notes.
% Additional author notes should be indicated with \thanks{} (for
% example, for current addresses).

% Example: \authors{A. B. Author\affil{1}\thanks{Current address, Antartica}, B. C. Author\affil{2,3}, and D. E.
% Author\affil{3,4}\thanks{Also funded by Monsanto.}}

\authors{F. E. Yan\affil{1}, H. Frezat\affil{2}, J. {Le Sommer}\affil{3}, J. Mak\affil{1,4}, K. Otness\affil{6}}

% \affiliation{1}{First Affiliation}
% \affiliation{2}{Second Affiliation}
% \affiliation{3}{Third Affiliation}
% \affiliation{4}{Fourth Affiliation}

\affiliation{1}{Department of Ocean Science, Hong Kong University of Science and Technology}
\affiliation{2}{Institut de Physique du Globe de Paris}
\affiliation{3}{Universit\'e Grenoble-Alpes, CNRS UMR IGE}
\affiliation{4}{National Oceanography Centre, Southampton}
\affiliation{5}{Courant Institute of Mathematical Sciences, New York University}
%(repeat as many times as is necessary)

%% Corresponding Author:
% Corresponding author mailing address and e-mail address:

% (include name and email addresses of the corresponding author.  More
% than one corresponding author is allowed in this LaTeX file and for
% publication; but only one corresponding author is allowed in our
% editorial system.)

% Example: \correspondingauthor{First and Last Name}{email@address.edu}

\correspondingauthor{Fei Er Yan}{feyan@connect.ust.hk}
\correspondingauthor{Julian Mak}{julian.c.l.mak@googlemail.com}

%% Keypoints, final entry on title page.

%  List up to three key points (at least one is required)
%  Key Points summarize the main points and conclusions of the article
%  Each must be 140 characters or fewer with no special characters or punctuation and must be complete sentences

% Example:
% \begin{keypoints}
% \item	List up to three key points (at least one is required)
% \item	Key Points summarize the main points and conclusions of the article
% \item	Each must be 140 characters or fewer with no special characters or punctuation and must be complete sentences
% \end{keypoints}

\begin{keypoints}
\item Explores an adjoint-based online learning approach for baroclinic turbulence, motivated by ocean eddy parameterization applications
\item Considers both an online and an approximate online approach, with the not requiring a differentiable model
\item The full online learning leads to more skillful and numerically stable model integrations
\end{keypoints}

%% ------------------------------------------------------------------------ %%
%
%  ABSTRACT and PLAIN LANGUAGE SUMMARY
%
% A good Abstract will begin with a short description of the problem
% being addressed, briefly describe the new data or analyses, then
% briefly states the main conclusion(s) and how they are supported and
% uncertainties.

% The Plain Language Summary should be written for a broad audience,
% including journalists and the science-interested public, that will not have 
% a background in your field.
%
% A Plain Language Summary is required in GRL, JGR: Planets, JGR: Biogeosciences,
% JGR: Oceans, G-Cubed, Reviews of Geophysics, and JAMES.
% see http://sharingscience.agu.org/creating-plain-language-summary/)
%
%% ------------------------------------------------------------------------ %%

%% \begin{abstract} starts the second page

\begin{abstract}  % 250 words
For reasons of computational constraint, most global ocean circulation models used for Earth System Modeling still rely on parameterizations of sub-grid processes, and limitations in these parameterizations affect the modeled ocean circulation and impact on predictive skill. An increasingly popular approach is to leverage machine learning approaches for parameterizations, regressing for a map between the resolved state and missing feedbacks in a fluid system as a supervised learning task. However, the learning is often performed in an `offline' fashion, without involving the underlying fluid dynamical model during the training stage. Here, we explore the `online' approach that involves the fluid dynamical model during the training stage for the learning of baroclinic turbulence and its parameterization, with reference to ocean eddy parameterization. Two online approaches are considered: a full adjoint-based online approach, related to traditional adjoint optimization approaches that require a `differentiable' dynamical model, and an approximately online approach that approximates the adjoint calculation and does not require a differentiable dynamical model. The online approaches are found to be generally more skillful and numerically stable than offline approaches. Others details relating to online training, such as window size, machine learning model set up and designs of the loss functions are detailed to aid in further explorations of the online training methodology for Earth System Modeling.

\end{abstract}

\section*{Plain Language Summary}  % 200 words

Most global ocean circulation models used in Earth System Modeling are still deficient in representing small-scale processes, affecting the modeled ocean circulation and its trustworthiness for making predictions. An increasingly popular approach is to use machine learning to represent the small-scale feedbacks. However, the learning is often performed in an `offline' fashion, without explicitly involving the underlying fluid dynamical model that may provide additional constraints and information to aid in model skill and stability. Here, we explore two `online' approaches, both of which involves the fluid dynamical model during the training stage for the problem of ocean turbulence. The full online approach is found to be generally more skillful and numerically stable. An approximately online approach, requiring substantially less initial investment in the model development, is found to inherit some but not all desirable properties of the full online approach.

%% ------------------------------------------------------------------------ %%
%
%  TEXT
%
%% ------------------------------------------------------------------------ %%

%%% Suggested section heads:
% \section{Introduction}
%
% The main text should start with an introduction. Except for short
% manuscripts (such as comments and replies), the text should be divided
% into sections, each with its own heading.

% Headings should be sentence fragments and do not begin with a
% lowercase letter or number. Examples of good headings are:

% \section{Materials and Methods}
% Here is text on Materials and Methods.
%
% \subsection{A descriptive heading about methods}
% More about Methods.
%
% \section{Data} (Or section title might be a descriptive heading about data)
%
% \section{Results} (Or section title might be a descriptive heading about the
% results)
%
% \section{Conclusions}

\section{Introduction}\label{intro}

Earth System Models and their component models are invaluable tools that enable us to perform numerical experiments, to probe how the Earth system might respond under various scenarios, to aid in our fundamental understanding of the Earth system itself, and to provide us with information that guide various mitigation strategies as the system evolves \cite<e.g.,>{IPCC-AR6}. Yet the various components of Earth System Models possess deficiencies, impacting the fidelity of the models in known and unknown ways. The present work focuses on the representation of physical processes in fluid systems, specifically on baroclinic turbulence. While the principal motivation for the present work is on ocean modelling, baroclinic turbulence is a generic feature in rotating stratified fluid systems, so the work here will have some physical relevance for atmospheric modelling. The numerical methodology should however be of general interest to any component of the Earth System model that possesses a mathematical description of the associated processes, particularly if they are described by partial differential equations.

Fluid systems such as the ocean or the atmosphere are multi-scale in nature, and the explicit representation of the their mechanical or thermodynamical processes depend on the choice of computational grid resolution. Ultimately there is a constraint on computational resources, so generally there will be sub-grid processes that lack an explicit representation in those models. The representation of these sub-grid processes and their feedbacks onto the resolved state, i.e. \emph{parameterizations}, critically impact the performance of the models. There have been increasing use of machine learning techniques to supplement and/or replace existing parameterizations, in the atmospheric context \cite<e.g.,>{BrenowitzBretherton19, YuvalOGorman20, Kashinath-et-al21, Mooers-et-al21, LopezGomez-et-al22, Connolly-et-al23, Sun-et-al23, Kochkov-et-al24} as well as the oceanic context \cite<e.g.,>{BoltonZanna19, ZannaBolton20, GuillauminZanna21, Frezat-et-al22, Ouala-et-al23, Perezhogin-et-al23, Ross-et-al23, Sane-et-al23, Yan-et-al23, Srinivasan-et-al24}. A standard approach in most of these works is to take a high resolution model and/or observational data (e.g.,satellite data) as the `truth', which is then filtered or coarse-grained in some way. The goal is then to obtain a mapping from a machine learning algorithm, where the input would normally be the filtered data, and the output or predicted quantity would be some residual term that arises from the choice of filtering (e.g., the $S^q$ quantity in the present work for a spatial average, or the eddy fluxes for a Reynolds average). The skill of the resulting mapping might then be judged on its ability to predict the withheld data that was not exposed to the model during the training stage \cite<e.g.,>{BoltonZanna19, GuillauminZanna21, Yan-et-al23}, referred to as \emph{a priori testing}. The aim is then to implement that mapping into a coarse resolution model with reduced computational demands, and the coarse resolution model is evolved in time with a state-dependent forcing as predicted by the machine learning mapping. There is of course no guarantee the resulting dynamical model has any skill or is even stable in this arguably more relevant \emph{a posteori test} \cite<e.g.,>{ZannaBolton20, GuillauminZanna21, Frezat-et-al22, ChattopadhyayHassanzadeh23, Ross-et-al23, Srinivasan-et-al24}. Some ways to stabilize the model integrations are often required or desirable, such as improving on the numerics associated with the inherent issues present machine learning algorithms \cite<e.g.,>{ChattopadhyayHassanzadeh23}, or forcing conservation properties \cite<e.g.,>{BoltonZanna19, Ross-et-al23}. 

We note that the aforementioned training procedure does not explicitly involve the dynamical model at the training stage, and might be termed an `offline' learning approach; the training is regarded as a regression between inputs and outputs that have no temporal correlation/causality explicitly invoked. Involving the dynamical model during the training stage over some time window in this `online' learning approach would be expected to lead to multiple benefits in the resulting model skill and/or stability, by forcing the training procedure to (1) recognize that the provided data has a causality/correlation as dictated by the dynamical model, (2) sample some of the statistics of the solution as encoded by the integration of the dynamical model, (3) exclude the set of possible mappings that lead to unstable model integrations, and possibly other desirable features. There are many possible methodologies available, with the crucial limitation being computational cost, usually related to the number of model parameters (e.g., number of degrees of freedom in a Convolutional Neural Network) and number of model evaluations involved (e.g., cost of the dynamical model run). Derivative-free methodologies tend to require more model evaluations and has a larger complexity scaling with the number of model parameters: a notable example is the Ensemble Kalman Inversion approach \cite<e.g.,>{Iglesias-et-al13, KovachkiStuart19, LopezGomez-et-al22, Schneider-et-al22}, although the use of emulations can substantially reduce the cost per model evaluation, and the methodology has some other desirable properties \cite<e.g.,>{Iglesias-et-al13}. On the other hand, derivative-based methods such as adjoint optimization has a lower complexity scaling with number of model parameters \cite<linear complexity scaling for adjoint-based methodologies; e.g.,>{Gunzburger-control}, making them particularly attractive for problems with a large number of parameters (e.g., for Convolution Neural Networks being considered in this work), although it does require a `differentiable' model, to be elaborated on shortly. The adjoint-based online approach is the one we consider in this work. The more classical adjoint approach has had important applications in data assimilation \cite<e.g.,>{Kalnay-DA} and in oceanography \cite<e.g.,>{Forget-et-al15a, Heimbach-et-al19}. The adjoint-based online learning approach has been shown to lead to significant improvements in fluid simulations \cite<called `temporal unrolling' in the works of>{List-et-al22, List-et-al24}. In studies of idealized but ocean-relevant studies, the works of \citeA{Frezat-et-al22} and \citeA{Ouala-et-al23} demonstrate that the adjoint-based online learning approach can lead to substantial improvements in the simulation of rotating turbulence in a one-layer system. A particularly impressive demonstration of improvements to model skill in atmospheric modeling with adjoint-based online learning is reported in the work of \citeA{Kochkov-et-al24}: the resulting neural general circulation model keeps the underlying atmospheric dynamical core, but achieves substantial speed up from replacing certain model processes by machine learned mappings, with notable improvements in weather and some aspects of climate forecasts over existing operational models.

A notable restriction with deploying adjoint-based online learning is the need for a `differentiable' model \cite<e.g.,>{Frezat-et-al22, Gelbrecht-et-al23, Sapienza-et-al24}. The problem is analogous to that of deriving an adjoint model in the applications of adjoint optimization \cite<e.g.,>{Forget-et-al15a, Heimbach-et-al19}: the adjoint associated with the dynamical model is required for performing the back-propagation calculations to compute the derivative of the loss function with respect to the parameters. For this, we either derive the adjoint associated with the dynamical model by hand and then discretize, or we discretize the dynamical model and then derive the numerical adjoint \cite<e.g.,>{Maddison-et-al19}. The former effectively requires maintaining two numerical models at the same time, and presumably any changes to one needs to be consistently changed in the other. The latter is in principle possible with algorithmic differentiation capabilities \cite<e.g.,>{Gelbrecht-et-al23, Sapienza-et-al24}. However, neither of these choices are normally taken at the initial stage of dynamical model development (with notable exceptions, e.g., MITgcm for adjoint optimization), and adjoint-based online learning as intended is in general not possible, not because of the machine learned mappings, but because the dynamical model does not by default have an associated adjoint. The previous works on online learning essentially wrote custom codes for idealized models \cite<e.g.,>{Frezat-et-al22, Ouala-et-al23, List-et-al22, List-et-al24} or rewriting an operational dynamical core \cite{Kochkov-et-al24} with algorithmic differentiation capabilities, enabling the adjoint-based online learning approach as intended to be performed. On that note, however, there has also been a recent proposals to alleviate the need of a differentiable dynamical model by approximating some of the calculations accordingly \cite{Ouala-et-al23, List-et-al24}. These approximately online approaches are of particular interest and applicability, since it may inherit some of the benefits of the full online learning and alleviate the need of a differentiable model, although the ideas and proposals require further testing and exploration.

Guided by the recent results and proposals for online learning, a principal aim of the present work is to provide an evaluation of the possible benefits and limitations of online learning over offline learning approaches for baroclinic turbulence. The evaluation of the different methodologies will utilize to a large extent an established benchmark procedure in an ocean-motivated setting by \citeA{Ross-et-al23}. In Sec.~\ref{methodology} we provide an overview of the methodology, detailing the differences between offline and online learning, the latter in the full form with an adjoint, and an approximate online form proposed by \citeA{List-et-al24}. The benchmark set up, as well as details of the differentiable dynamical model with algorithmic differentiation capabilities are detailed as part of the methodology. Sec.~\ref{results_eddy} provides a comprehensive evaluation of the results from the different methodologies. Sec.~\ref{online} provides additional details procedures that are more specific to online learning. Sec.~\ref{conclusion} summarizes the results of the work and provides an outlook to further usage of online learning in the Earth System Modeling setting.

%%%%%%%%%%%%%%%%%%%%%%%%%%%%%%%%%%%%%%%%%%%%%%

\section{Methodology} \label{methodology}

In this section we first detail what we mean by offline and online learning, highlighting similarities and links in the principles and nomenclature between the Earth System Science, machine learning, and classical optimization. We will opt for using the terminology in classical optimization \cite<e.g.,>{Gunzburger-control} in our expository part. The details benchmark setting and the numerical model largely based on \citeA{Ross-et-al23} are then documented.

%-------------------------------------------------------
\subsection{Offline vs. online learning}

At its core, the problem we are interested in is one of regression: there is a high resolution model truth, and a low resolution model possibly with some extra terms added in, and the aim is to choose the extra terms so that the low resolution model is closer to the high resolution in some metric(s) to be specified. The problem can then be phrased as one of optimization: the objective is to choose the magnitude, distribution and/or form of the extra terms (the \emph{control variables}), such that differences in the target data from the high resolution model and the predictions from the modified low resolution model in a relevant metric, encoded in a \emph{loss function} (or sometimes cost, objective or mismatch function) is minimized. The extra terms and the loss function are defined in terms of a \emph{state vector}, consisting of the variables provided by the model directly, or some functions of those variables (e.g., the velocity from a fluid model, or the transport computed from an integral of the velocity). For concreteness, we will be referring to Convolution Neural Networks (CNNs) as the machine learning algorithm, although the exposition we provide is valid for other choices.

%%%

\subsubsection{Offline learning}

For the present focus on parameterization, one way to form the machine learning problem would be \emph{offline training} \cite<e.g.,>{Ross-et-al23}, or sometimes \emph{one-step learning} in the machine learning literature \cite<e.g.,>{List-et-al24}. Training data is obtained for example by filtering data from a high resolution model onto a low resolution grid, forming our state vector $\mathbf{y}$. A CNN denoted $\mathcal{N}(\mathbf{x}|\boldsymbol{\theta})$ is our mapping function from some input $\textbf{x}$ to the extra terms that we want to include into a low resolution model; the overbar denotes a filtering procedure, and the CNN weights $\boldsymbol{\theta}$ are the control variables. We aim to find the $\boldsymbol{\theta}$ that minimizes a loss function, usually defined as the mean squared error between the output of the CNN and the target data, i.e., we seek $\boldsymbol{\theta}$ such that
\begin{linenomath*}
\begin{equation}\label{eq:offline_lf}
  \mathcal{L}_{\rm off}= |\overline{\mathbf{y}} - \hat{\overline{\mathbf{y}}}|^2, \qquad \textnormal{with} \qquad \hat{\overline{\mathbf{y}}} = \mathcal{N}(\overline{\mathbf{x}}|\boldsymbol{\theta}),
\end{equation}
\end{linenomath*}
is minimized. The crucial observation here is that the construction of the loss function is defined without any reference to the dynamical model itself: it is essentially seeking the optimal map between the target data $\overline{\mathbf{y}}$ and the prediction $\hat{\overline{\mathbf{y}}} = \mathcal{N}(\overline{\mathbf{x}}|\boldsymbol{\theta})$. When a time-series of spatially varying data is provided, one should regard the data as multi-dimensional arrays that are appropriately reshaped or flattened into a collection of data points, and the optimal mapping $\mathcal{N}(\cdot|\boldsymbol{\theta})$ over the set of training data is the returned object.

In that sense, however, the temporal information is absent in offline learning, which is potentially problematic for data with a temporal dimension: causality in time provides additional constraints that may not be respected by the resulting offline models. While such offline models is likely to have skill in the mapping of the data at fix instances in time (i.e. \emph{a priori} testing), there is usually no guarantee that the hybrid dynamical models with the offline model implemented has skill or be numerically stable over time (i.e. \emph{a posteriori} testing). The numerical stability aspect is particularly problematic, and variable stabilizers may be required, ideally physically informed. The proposal of \citeA{Ross-et-al23} imposes a zero net potential vorticity tendency as a hard constraint in the CNNs trained offline, since eddy parameterization should redistribute but not create addition potential vorticity in the system; we have found that omitting such a hard constraint in our calculations can (but not always) lead to unstable model integrations (not shown). There are also other methodologies to enforce conservation, some of which will be discussed in Sec.~\ref{choix-loss}.

%%%

\subsubsection{Full (adjoint-based) online learning}

One might suspect including the dynamical model into the definition of the cost function and/or the training procedure might be beneficial. Instead of looking for optimal choices of control variables $\boldsymbol{\theta}$ over isolated instances in time, we might be more interested in the optimal choice over a solution trajectory defined over a time window. One reason is presumably that the inclusion of the dynamical model creates a mapping that forces the optimization to take into account that there is inherent temporal correlation/causality in the data. Further, the resulting optimization for the control variables is expected to remove the choices that lead to large deviations (e.g., associated with unstable model integrations in time), since that leads to growth in the loss function defined over a time window. The inclusion of the underlying dynamical model in the training and the training procedure over a time window are what we consider to be the key ingredients to \emph{online learning} \cite<e.g.,>{Frezat-et-al22, Ouala-et-al23, Kochkov-et-al24}; this was called \emph{temporal unrolling} in the machine learning literature \cite<e.g.,>{List-et-al22, List-et-al24}.

We make precise what we mean by an adjoint-based online learning in its `full' form first. Using the same terminology as above, online learning considers a loss function defined \emph{over a solution trajectory}, i.e.,
\begin{linenomath*}
\begin{equation}\label{eq:online_lf}
  \mathcal{L}_{\rm on}= \sum^K_{i=0} |{\overline{\mathbf{y}}}_{t_0+i\Delta t} - \hat{\overline{\mathbf{y}}}_{t_0+i\Delta t}|^2, \qquad \textnormal{with} \qquad \hat{\overline{\mathbf{y}}}_{t_0+i\Delta t} = M^i(\overline{\mathbf{x}}_{t_0}|\boldsymbol{\theta}),
\end{equation}
\end{linenomath*}
where $\mathbf{y}_{t_0+i\Delta t}$ is the state vector at time $t_0+i\Delta t$, $\Delta t$ is the (fixed) discretized time step, $\hat{\overline{\mathbf{y}}}_{t_0+i\Delta t}$ denotes the prediction of the state vector at the same time, and $K\Delta t$ is the extent of the time window (so $K$ is the number of time steps that defines the time window of training). The mapping $M$ is a hybrid dynamical model, i.e., a low resolution dynamical model with terms predicted by the CNNs, which generates the predicted state vector $\hat{\overline{\mathbf{y}}}_{t_0+i\Delta t}$ given some initial data denoted $\overline{\mathbf{x}}_{t_0}$. The objective is again to choose the CNN weights $\boldsymbol{\theta}$ (the control variable) such that $\mathcal{L}_{\rm on}$ is minimized. What has been mentioned so far would be what we would regard as `online' learning; the choice we make in employing an adjoint-based methodology utilizing the derivative information for the optimization problem makes it an adjoint-based online learning. One notable difference for online learning relative to offline approaches is that offline approach regards the whole set of data at the same time, while the former does not. Crudely, offline learning is akin to finding a curve/surface of best fit given the whole collection of data embedded in some high dimensional data space, while online learning is aiming to find the best fit \emph{trajectory} over the analogous high dimensional data space, with the time dimension having a special status. The approach is identical in spirit to that of Four-Dimensional Variational Assimilation \cite<4DVAR; e.g.,>{Kalnay-DA}, where the mapping could instead be the full dynamical model, the target data would be the observations of the atmosphere/ocean state (or related derived observables, e.g., the volume transport), and the control variable might be the initial state or other model parameters (e.g.,free parameters in existing parameterizations).

If we assume appropriate differentiability of $\mathcal{L}_{\rm on}$, we could solve the problem of\footnote{Abusing notation somewhat here, since $\boldsymbol{\theta}$ is formally a function, so we should be talking about Fr\'echet or variational derivatives.} $\partial \mathcal{L}_{\rm on} / \partial \boldsymbol{\theta} = 0$ to find the (local) minimum of $\mathcal{L}_{\rm on}$, which may be done with an adjoint associated with the mapping $M$ given the choice of $\mathcal{L}_{\rm on}$. Adjoint-based methods use $M$ to perform a forward sweep leading to an evaluation of $\mathcal{L}_{\rm on}$, and the associated adjoint performs a back-propagation calculation that leads to an evaluation of $\partial \mathcal{L}_{\rm on} / \partial \boldsymbol{\theta}$ via the chain rule \cite<e.g.,>{Gunzburger-control, List-et-al22, Ouala-et-al23, List-et-al24}. A technical requirement for adjoint-based methods is that $M$ needs to differentiable, in the sense that one can derive the associated adjoint, by hand or algorithmically. The CNN part is not a problem since these are combinations of maps and activation functions implemented in software that enables automatic differentiation capabilities, but most dynamical models are not constructed to have a manual adjoint or have the capabilities to generate an adjoint automatically via algorithmic differentiation \cite<e.g.,>{Ouala-et-al23, List-et-al24, Sapienza-et-al24}. 

If the dynamical model is differentiable, then the back-propagation of the sensitivities for the gradient calculations can be done throughout the whole time window of interest, and this might be called `end-to-end', in that the information can be propagated from one end to another in both directions \cite<e.g.,>{Ouala-et-al23, List-et-al24, Sapienza-et-al24}. Note that just as the dynamical model can have issues with numerical stability in the forward or prediction mode, the associated adjoint model can also have numerical instabilities for the back-propagation calculations, resulting in numerical issues with the gradients $\partial \mathcal{L}_{\rm on} / \partial \boldsymbol{\theta}$ and the update stage involving the iterative solver for the next forward-backward sweep. Some manual approaches are possible to force the back-propagation stage to be more stable \cite<e.g., approaches detailed in>{List-et-al22}, although one could argue there is loss of consistency as a result.

%%%

\subsubsection{`Approximate' online learning}

While the ideal situation would be that we have access to a differentiable dynamical model with no extra stabilizers needed on the forward or back-propagation calculations, the fact is that most dynamical models are not differentiable, nor will the back-propagation step necessarily be numerically stable. One might however still want to perform a training procedure as detailed above to access some of the desirable features that online learning is expected to offer. One possibility is to use a methodology that does not require derivative information, in exchange for possible increases in computational complexity \cite<e.g.,>{Iglesias-et-al13, KovachkiStuart19, LopezGomez-et-al22, Schneider-et-al22}. Another is to approximate the relevant derivatives in the adjoint-based approach. The approach of \citeA{Ouala-et-al23} considers a first-order approximation to $\partial M / \partial \boldsymbol{\theta}$ (which arises from application of the chain rule to get the derivative of $\mathcal{L}_{\rm on}$), which they term the Euler Gradient Approximation, which they show converges back to the full adjoint-based approach in the limit of infinitesimal time-step size. The approach of \citeA{List-et-al24} considers the following pipeline: proceed as detailed above, but at each time-step, the back-propagation of the information only reaches the CNN stage, and is then set to zero. The resulting approximation is more drastic and heavily truncates the back-propagation calculation, but is particularly easy to implement. In that regard, both the aforementioned approximations target the back-propagation part, but otherwise solves the analogous optimization problem. One might suspect such approximations might be reasonable for short time windows \cite<or perhaps there are better approximation procedures available, as alluded to in>{Ouala-et-al23}. For the present work, we consider the approach of \citeA{List-et-al24} for simplicity. While we might suspect that the fully differentiable online procedure is most theoretically consistent and might be most skillful, it could also be that the approximated online procedures are in fact good enough and inherit some of the desirable properties of the full online procedure \cite{Ouala-et-al23, List-et-al24}. A secondary aim of the present work is to explore to what extent the approximate online procedure is comparable to the full online procedure.

%-------------------------------------------------------
\subsection{Experiment and model details} \label{model}

Building on the previous works of \citeA{Frezat-et-al22} and \citeA{Ouala-et-al23} in a one-layer quasi-geostrophic model, we utilize an idealized two-layer quasi-geostrophic turbulence model that allows for baroclinic instability, following the work of \citeA{Ross-et-al23}, who demonstrated their work on offline training in the \texttt{pyqg} model. To enable the full adjoint-based online approach to be performed, we utilize a model called \texttt{pyqg-JAX} \cite{Otness-et-al23, Otness24}, which has capabilities to generate the associated adjoint and perform full back-propagation through the dynamical model via the inbuilt algorithmic differentiation capabilities. We utilize the same model settings as in the previous work of \citeA{Ross-et-al23} in the `eddy' regime. Briefly, the model is formulated on a doubly periodic $\beta$-plane, governed by
\begin{linenomath*}
\begin{align}
  \partial_t q_1+\nabla \cdot (\mathbf{u}_1 q_1)+\beta \partial_x \psi_{1} + U_1 \partial_x q_{1} &= \mathrm{ssd}, \label{eq:2-layer qg 1} \\
  \partial_t q_2+\nabla \cdot (\mathbf{u}_2 q_2)+\beta \partial_x \psi_{2} + U_2 \partial_x q_{2} &=-r_{ek}\nabla^2\psi_2 + \mathrm{ssd}, \label{eq:2-layer qg 2}
\end{align}
\end{linenomath*}
where subscripts 1 and 2 denote the upper and lower layer, and $\partial_{x,y}$ denote the derivatives in the $x$ and $y$ (or zonal and meridional) directions. The mean zonal velocities $U_1$ and $U_2$ are set to maintain a prescribed vertical shear that drives the turbulence in the model. Denoting $l$ as the layer index, the perturbation variables are the potential vorticity $q_l = \nabla^2 \psi_l + \beta y + (-1)^l f_0^2/(g' H_l) (\psi_1 - \psi_2)$, $f_0$ the planetary Coriolis parameter, $\beta$ is the meridional variation of the planetary rotation, $g'$ the reduced gravity measure relating to the stratification between layer 1 and 2 with depth $H_{1,2}$, the streamfunction $\psi_l$, with the geostrohic velocity given by $\mathbf{u} = (u, v) = (-\partial_y \psi, \partial_x \psi)$, $r_{ek}$ is the Ekman friction acting only on the bottom later dissipating large-scale energy, and $\mathrm{ssd}$ denotes the small-scale dissipation, which dissipates small-scale energy and potential enstrophy $q^2$. The relevant parameters of the model are summarized in Table~\ref{tbn:pyqg}.

\begin{table}[tb]
  \begin{center}
  {\small
    \begin{tabular}{c c c}
      \hline
      parameter & symbol & value and units\\
      \hline
      Domain size & $L\times L$ & $1000\ \mathrm{km} \times 1000\ \mathrm{km}$\\
      Meridional planetary vorticity gradient &$\beta$ & $1.5\times10^{-11}\ \mathrm{m}^{-1}\ \mathrm{s}^{-1}$\\
      Bottom friction coefficient &$r_{\rm ek}$ & $5.787\times10^{-7}\ \mathrm{s}^{-1}$\\
      Layer thickness &$(H_1, H_2)$ & $(500, 2500)\ \mathrm{m}$\\
      Rossby deformation radii &$ L_d $ & $ 15\ \mathrm{km}$\\
      prescribed mean zonal flows &$(U_1, U_2)$ & $(0.025, 0)\ \mathrm{m/s}$\\
    
      \hline
    \end{tabular}
  }
  \end{center}
  \caption{Summary of parameters used for the two-layer quasi-geostrophic model.}
  \label{tbn:pyqg}
\end{table}

The model is discretized in the pseudo-spectral formalism, with the derivatives computed in Fourier spectral space while the multiplication is performed in real space, utilizing a 2/3 de-aliasing procedure \cite<e.g.,>{Orszag71b}. The small-scale dissipation $\mathrm{ssd}$ is implemented as a quadruple-exponential filter, which acts as a low-pass filter and attenuates wavenumbers above a cutoff threshold $\kappa_c$, which is 2/3 of the model's Nyquist frequency $\kappa_{N_y} = \pi$, i.e.
\begin{linenomath*}
  \begin{equation}\label{eq:sharp_filter}
   q_k = \left\{ \begin{array}{lcl} q_k  & \mbox{for} & \kappa  < \kappa_c, \\ 
                     q_k \mathrm{e}^{-23.6(\kappa -\kappa_c)^4} & \mbox{for} & \kappa \ge\kappa_c,
                  \end{array}\right.
  \end{equation}
\end{linenomath*}
where $q_k$ is the Fourier coefficient of $q$ in spectral space. The value of the e-folding scale $-23.6$ is chosen so that the filtering is numerically zero at the largest wavenumber of the system. The model is discretized in time with a third-order Adams--Bashforth scheme.

\begin{figure}[tb]
\begin{center}
	\includegraphics[width=1.0\textwidth]{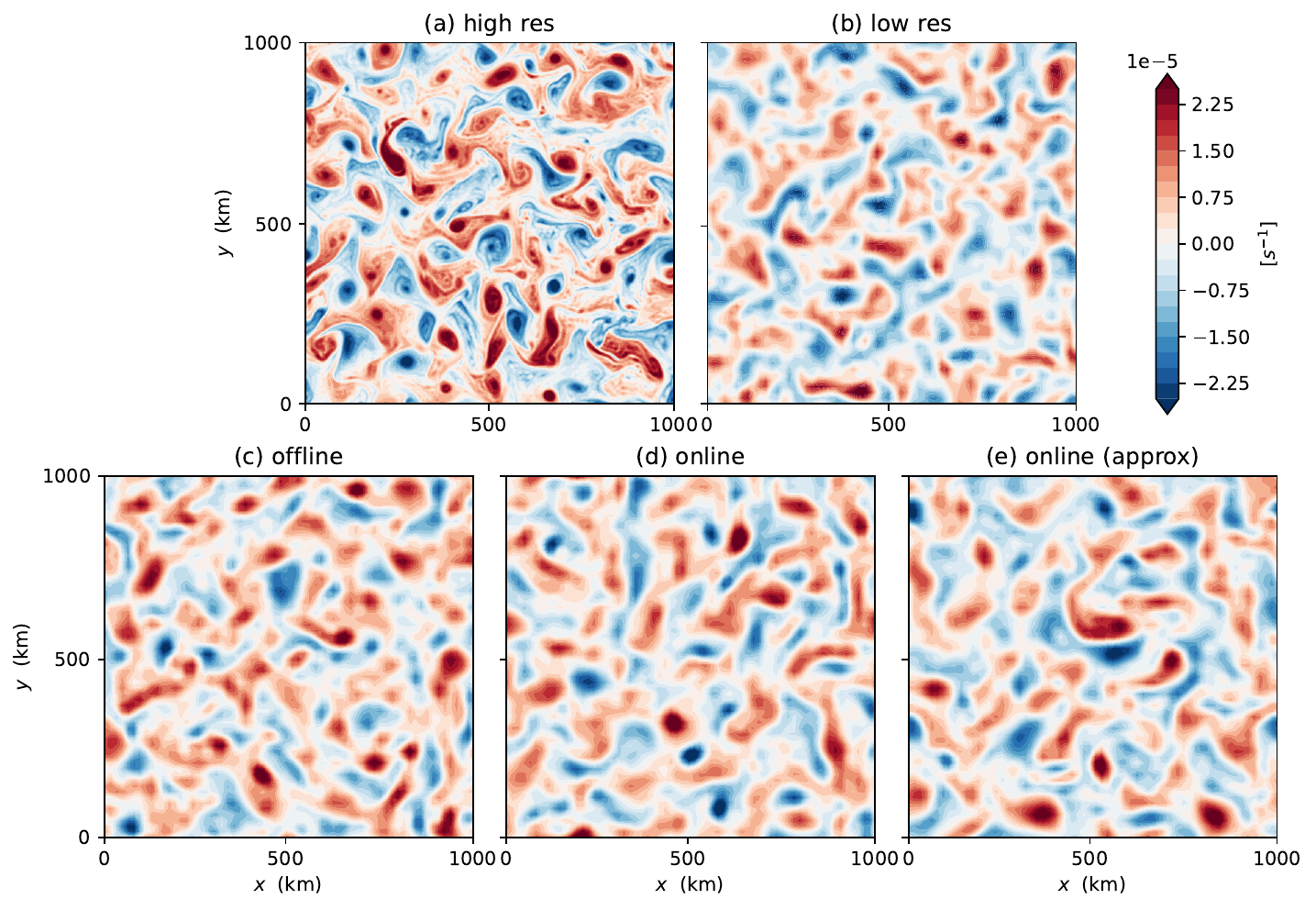}
	\caption{Snapshots of upper-layer perturbation potential vorticity $q_1$ at the end of a ten year simulation. Data from the ($a$) high resolution model true, ($b$) low resolution model with no CNN active, ($c$) offline model, ($d$) online model ($e$) approximately online model.}
	\label{fig:results_eddy_q}
\end{center}
\end{figure}

As discussed in \citeA{Hallberg13}, the model grid size $\Delta x$ $(= \Delta y)$ should be smaller than the deformation radius $L_d=15\ \mathrm{km}$, and ideally we should have $L_d/\Delta x \ge 2$ to resolve the mesoscale eddies. We use a model grid with size $256^2$, corresponding to $\Delta x = 3.9\ \mathrm{km}$ (with a time-step of $\Delta t = 1$ hour) as a reference high resolution simulation. For the low resolution simulation, we take instead a grid size of $64^2$, corresponding to $\Delta x = 15.6 \mathrm{km}$ (with a time-step size of $\Delta t = 4$ hours). All models start with a randomly generated initial condition, and all the simulations are run for 10 model years to reach the equilibrium. Sample outputs of the upper layer perturbation potential vorticity $q_1$ are shown in Fig.~\ref{fig:results_eddy_q}$a,b$ for the high and low resolution calculations respectively. Note that the present choice of the bottom drag time-scale $r_{\rm ek}$ leads to an `eddy' rather than `jet' regime \cite<e.g.,>{Ross-et-al23}; we discuss the analogous results for the `jet' regime in Sec.~\ref{conclusion}.

%-------------------------------------------------------
\subsection{Training strategy} \label{training}

For both offline and online approaches, we follow the strategy of large eddy simulation and the processing as in \citeA{Ross-et-al23}. We first coarse-grain Eq.~(\ref{eq:2-layer qg 1}) and (\ref{eq:2-layer qg 2}) to get the filtered equations; in this work we coarse-grain by a spectral truncation, since we already utilize a pseudo-spectral formalism. We take the state vector $\boldsymbol{y}$ to be the sub-grid forcing $S^q$, which are the resulting terms in the coarse-grained equations, given by
\begin{linenomath*}
  \begin{equation}\label{eq:sq0}
    S^q=\overline{ \nabla \cdot (\mathbf{u} q) }- \nabla \cdot (\overline{\mathbf{u} q}) = \overline{(\mathbf{u} \cdot \nabla) q} - (\overline{\mathbf{u}} \cdot \nabla)\overline{q}.
  \end{equation}
\end{linenomath*}
The overbar denotes the coarse-graining by the spectral truncation, and $S^q$ is the residual between the coarse-grained advective potential vorticity tendency and the advective potential vorticity tendency formed from the coarse-grained fields. The input for all training strategies is chosen to be the potential vorticity $q$, obtained either by filtering the high resolution model model for \emph{a priori} testing, or the $q$ field as simulated by the low resolution hybrid models for \emph{a posteriori} testing. The task of the machine learning algorithms is to predict the missing tendency $S^q$, and the models differ only in how the control variables (i.e. the CNN weights) are adjusted during the training phase. Samples of the upper layer perturbation potential vorticity $q_1$ from the hybrid models (low resolution model with the corresponding machine learning model implemented) for a prognostic prediction are shown in Fig.~\ref{fig:results_eddy_q}$c,d,e$; we note the resulting $q$ field from the hybrid models with a CNN are marginally sharper than the low resolution model with no CNN active (Fig.~\ref{fig:results_eddy_q}$b$).

For systems such as the present multi-layer quasi-geostrophic model, there is a choice on how one chooses the loss function. One choice is to construct one single loss function $\mathcal{L}$ based on a weighted sum of the loss function computed from the separate layers $\mathcal{L}_{1,2}$, i.e.
\begin{linenomath*}
  \begin{equation}\label{eq:lf_1loss}
    \mathcal{L}= \mathcal{L}_1+ \alpha \mathcal{L}_2,
  \end{equation}
\end{linenomath*}
where, for this work, the layer-wise loss function is defined as
\begin{linenomath*}
\begin{equation}\label{eq:Ll}
  \mathcal{L}_{1,2} = \sum^K_{i=0} |S^q_{1,2;\ t_0+i\Delta t} - \hat{S}^q_{1,2;\ t_0+i\Delta t}|^2,
\end{equation}
\end{linenomath*}
where $\hat{S}^q_{1,2}$ denotes the predictions from the CNN; for the offline approach, we formally take $K=1$ but consider the $S^q$ field as a stacked array. The parameter $\alpha$ is introduced to provide a balance so that the data between the layers have comparable magnitudes (e.g., $S^q$ in layer 1 in this model is at least two orders of magnitude layer than that in layer 2), or as our prior bias in which layer we favor. While one could argue the resulting model will be exposed to a wider regime of data by sampling more of the probability distribution, the model could also be argued to be less specialized, and there is subjectivity in how the free parameter $\alpha$ is chosen. Another approach is simply to train up separate CNNs via separate loss function $\mathcal{L}_{1,2}$, such as that following the work of \citeA{Ross-et-al23}; two CNNs are thus at play, costing more in terms of training and prediction. We take the latter approach in this work, but discuss the performance of combined or separate choice of loss function(s) in Sec.~\ref{choix-loss}.

Note that it is in fact possible to use the state variables $q$ or $\psi$ directly for online learning, as was done in a previous work of a one-layer model \cite{Frezat-et-al22, Ouala-et-al23}. The difference is that instead of targeting the sub-grid tendencies (which are small-scale quantities), it is possible to target the mean-state variables directly. For the present work we were interested in providing a like-for-like comparison between the methodologies as well as resulting from previous works \cite<e.g.,>{Ross-et-al23}, so training with loss functions defined in terms of $q$ or $\psi$ was not pursued in detail; we present some related results in Sec.~\ref{choix-loss} and comment on the procedure in Sec.~\ref{conclusion}.

A technical issue is that \texttt{pyqg-JAX} assumes data uses single-precision floats by default, but the diagnosed magnitude of $S^q$ in both layers are small, roughly $10^{-11}$ in layer 1 and $10^{-13}$ in layer 2. In the present work we employ the single-precision version, and the only standardization we perform is multiplying $S^q$ and perturbation potential vorticity $q$ by $10^{13}$ and $10^6$ in both layers, with the relevant scale factor are added or removed at the appropriate stages. Empirical testing for this work suggests the choice of data standardization is not crucial to the performance of the models.

For both offline and online learning, we employ CNNs as the mapping between input and outputs. For commutativity with the differentiable solver \texttt{pyqg-JAX}, we use the JAX library \texttt{Equinox} \cite{KidgerGarcia21}. We follow the work of \citeA{Ross-et-al23} on the choice of CNN architecture in \texttt{pyqg}: the CNNs have eight hidden convolutional layers with square kernels, first two layers with size 5 and in circular padding mode, and the remaining six layers have size 3 in constant padding mode. The stride for the cross-correlation is 2 for the first two layers, and 1 for other layers. Following \citeA{Ross-et-al23} and as mentioned earlier, a hard constraint is imposed so that there is no net tendency of potential vorticity added by the predicted $\hat{S}^q$, motivated by the observation that the nonlinear transfers being parameterized should transfer but not create or destroy potential vorticity, i.e.,
\begin{linenomath*}
\begin{equation}\label{eq:constraint}
  \int\hat{S}^q\; \mathrm{d}A = 0.
\end{equation}
\end{linenomath*}
There are at least three possible meanings here, that the training for the CNN weights $\boldsymbol{\theta}$: 1) results as usual with no regard of the constraint, and the resulting prediction has the mean removed \emph{post-hoc}; 2) takes into account that the prediction will need to pass through a de-meaning procedure as the final CNN layer; 3) is such that the choice of $\boldsymbol{\theta}$ is restricted such that the results prediction has zero mean. The second case is what is meant here. The first case is arguably \emph{ad hoc}, while the third case would be theoretically most satisfactory, but is likely difficult to execute in practice; a further discussion is given in Sec.~\ref{choix-loss}. ReLU is chosen as the activation function between each layer. No batch normalization is performed in JAX for simplicity.

For offline learning, we generate the training data by performing 50 high resolution simulations with different randomized initial conditions, and each of these are integrated forward in time for 10 model years (86,400 time steps). We sample the all high resolution simulations every 500 time-steps as a means to decorrelate data temporally, coarse-grain the data by spectral truncation, and randomize the ordering of the whole set of data frames, so the choice of random seed affects the partition of the data frames into the training-testing datasets. For online learning, the training data is a temporal sequence of data frames. A total of 10 high resolution simulations were performed, and the choice of random seed affects the exact time window selected over that whole possible set of trajectories from the high resolution model on which the CNN is trained on. The offline CNNs are trained with a batch size of 32 for 800 epochs in a NVIDIA Tesla T4 GPU in Google Colab. For online learning, we gradually increase the window size from 2 time-steps to 10 time-steps in strides of 2 for training stability, i.e. starting from a converged model and using that as the initial condition for the longer set of learning \cite<e.g.,>{Frezat-et-al22}; this is referred to as a `curriculum' in \citeA{List-et-al22}, and is related to continuation approaches in optimization (solving a related but easier and/or more stable problem, which gradually transforms into the desired problem). Note that the time window size is rather small, and further discussion on the choice of window size is given in Sec.~\ref{window_size}. The CNNs from both the online learning procedures are trained for 10 epochs at each temporal window size, and 50 epochs in total. For each of the training methodology (offline, online, and approximate online), we train an ensemble with ten members, each member within the set of ten differing only in the choice of the random seed (but the choice of seeds is uniform across all three ensembles). If the model calculation of an ensemble member fails at the training or the integration stage, we simply state the member is unstable and remove it from the model score calculations. 

We note that the offline procedure is exposed to more time frames of the data, but the online procedure has the data temporally correlated, and it is problematic to gauge whether the two strategies tested here are exposed to the same amount of information in the training data for a completely fair test. We have attempted to vary the amount of data that is exposed to the training strategies: we report on the tests with the online case later in Sec.~\ref{window_size}, but the results reported for offline training in this work are robust to increases in number of data frames provided at the training period.

%%%%%%%%%%%%%%%%%%%%%%%%%%%%%%%%%%%%%%%%%%%%%%%%%%%

\section{Results relating to model skill}\label{results_eddy}

To examine the model performance between offline and online learning, we perform both \emph{a priori} and \emph{a posteriori} testing. Recall from above that the CNNs expects the (perturbation) potential vorticity $q$ as the input field, and outputs a sub-grid forcing $\hat{S}^q$. \emph{A priori} tests for the model skill in being able to predict snapshots of diagnosed $S^q$ for given filtered high resolution potential vorticity field $q$, while \emph{a posteriori} testing will be related to model performance while the model is being integrated forward in time, for relevant model metrics to be defined. We note that for the present set of results, none of the models crashed during training or when integrated in time in prognostic mode for the choice of random seeds we have selected, and the results are reported for the full ensemble for offline, full online and approximately online approaches; the model stability seems to be a consequence of the imposed hard constraint given by Eq.~\eqref{eq:constraint}, which we further comment on in Sec.~\ref{choix-loss} and \ref{conclusion}.

%-------------------------------------------------------
\subsection{Distribution similarity}

\begin{figure}[tb]
\begin{center}
	\includegraphics[width=1.0\textwidth]{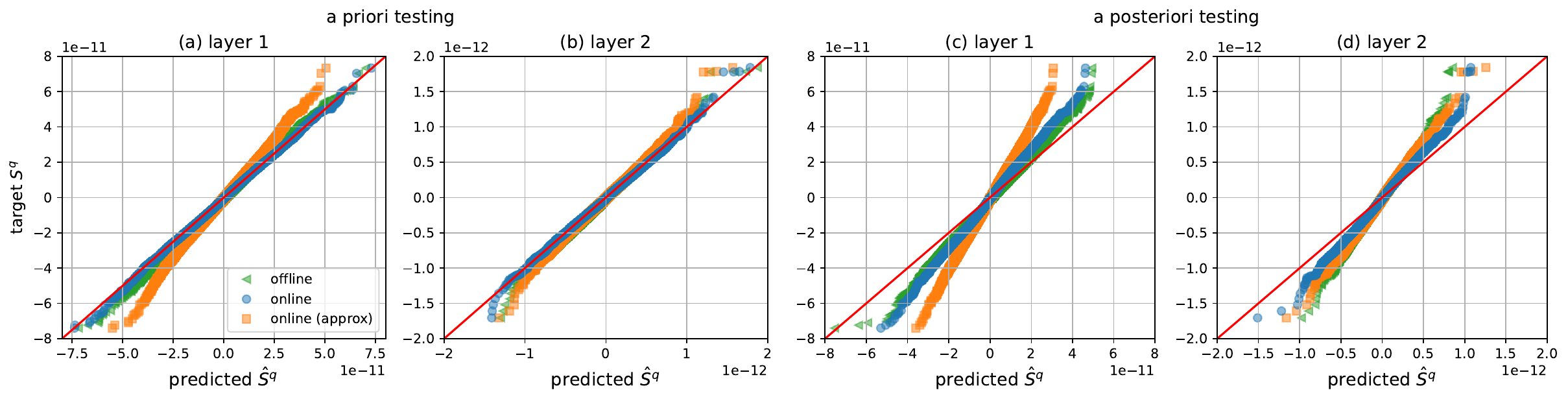}
	\caption{Quantile-Quantile (Q-Q) plot, with the distribution of diagnosed target (the $S^q$ from high resolution filtered onto the coarse grid) on the $x$ axis, against the predicted $\hat{S}^q$ generated by the CNNs on the $y$ axis (green: offline model; orange: online (approx) model; blue: online model). ($a,b$) Results from \emph{a priori} testing, and ($c,d$) \emph{a posteriori} testing, for layer 1 and 2 respectively. The red line is the identity line $y=x$.}
	\label{fig:results_qqplot}
\end{center}
\end{figure}

In Fig.~\ref{fig:results_qqplot} we show quantile-quantile (Q-Q) plots to assess the distribution similarities between the targets $x$ and the predictions $y$: dissimilarities in the quantiles arise as deviations from the $y=x$ line. We show results from the \emph{a priori} test for both fluid layers in Fig.~\ref{fig:results_qqplot}$a,b$, using the same group of input $q$ and output $S^q$ for the trained CNNs. Here it would appear that offline and the full online approach is able to generate predictions that largely agree in the distribution, while the approximate online approach has rather large biases, particularly in the top layer.

For the \emph{a posteriori} testing, the same initial state $q$ is provided to all hybrid models, and these models are integrated forward in time for 10 model years (86,400 time-steps), sampled every 20 time-steps. The predicted $\hat{S}^q$ from the hybrid models are compared to the diagnosed $S^q$ from the high resolution model, and the corresponding Q-Q plots are shown in Fig.~\ref{fig:results_qqplot}$c,d$. Since there is inherent data drift as the models run in prognostic mode, the deviations away from the $y=x$ line are larger under \emph{a posteriori} testing for all cases. Overall, it is the full online approach that leads to the best performance in terms of the distribution.

%-----------------------------------------------------------
\subsection{Energy budget}

\begin{figure}[tb]
\begin{center}
	\includegraphics[width=\textwidth]{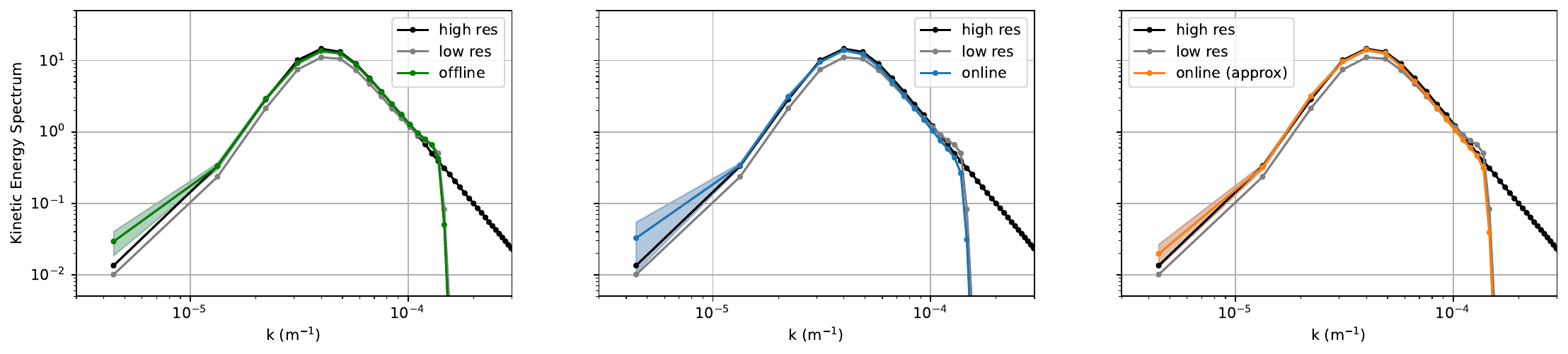}
	\caption{Depth-averaged total kinetic energy spectra, normalized by $n^2_x \times n^2_y$. All panels show the data from the target high resolution simulation (black line) and the low resolution simulation with no CNN active (grey line). Results from the ($a$) the offline model (green line), ($b$) full online model (blue line), and ($c$) the approximately online model (orange line). The colored lines are from a ten member ensemble averaged performance of the hybrid models, and the shading denotes the standard deviation of the ensemble.}
	\label{fig:results_spectral_ke}
\end{center}
\end{figure}

The accurate representation of energy levels is crucial in numerical ocean models, since energy measures the variance of the dynamical variables. We show in Fig.~\ref{fig:results_spectral_ke} the time and depth-averaged total kinetic energy (KE) spectra over the last five model years, where KE density is defined as $|\mathbf{u}|^2/2$, and the spectra is computed assuming horizontal isotropy. It can be seen that the low resolution model (without a CNN active; grey line) generally lacks KE across all scales, and there is the `kick-back' feature at the high wavenumbers, symptomatic of energy accumulation at the grid-scale and a sign of the model being under-resolved. The hybrid models on the other hand generally resolve the bias that there is a lack of KE across all scales. The offline model possesses a kick-back in the KE spectra; we note this feature is not something that can be `tuned' out by adjusting the training, and is really a robust feature of the resulting offline models. On the other hand, neither of the online models show this kick-back feature for the reported time window size of ten time-steps; see Sec.~\ref{window_size} for further related results. All models have a KE that is too large at the large-scales, which is consistent with the results reported in \citeA{Ross-et-al23} and \citeA{Perezhogin-et-al23} but for offline models. 

\begin{figure}[tb]
\begin{center}
	\includegraphics[width=1.0\textwidth]{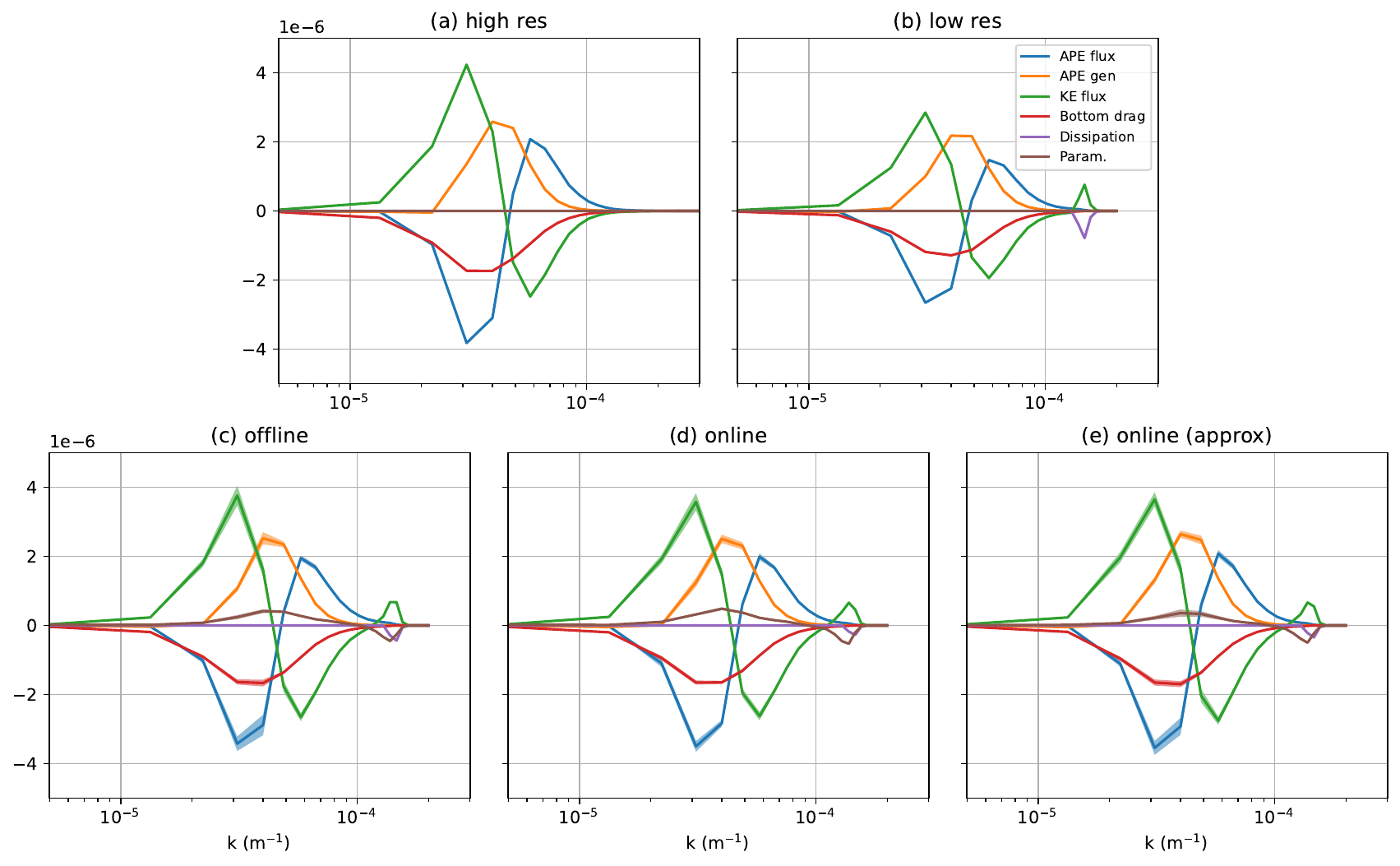}
	\caption{Full spectral energy budget for the APE flux (blue line), APE generation from the imposed background shear (orange line), KE flux (green line), bottom friction (red line), dissipation from the spectral filter (purple line), and input from the CNNs (brown line). ($a$) The high resolution model truth. ($b$) The low resolution model with no CNNs active. ($c$) The offline hybrid model. ($d$) The online hybrid model. ($e$) The approximately online hybrid model. For ($c,d,e$), the colored lines are from the ten member ensemble averaged performance of the hybrid models, where the shading denotes the standard deviation of the ensemble.}
	\label{fig:results_eddy_energy}
\end{center}
\end{figure}

While the KE spectra above tells us about the energy content distribution, it is informative to consider the energy fluxes. The full energy budget for the present idealized ocean model can be represented as \cite<e.g.>{JansenHeld14}
\begin{linenomath*}
  \begin{equation}\label{eq:energy budget}
    \partial_t \mbox{E} = \mbox{KE}_{\rm flux}+ \mbox{APE}_{\rm flux} + \mbox{APE}_{\rm gen} - F - D.
  \end{equation}
\end{linenomath*}
Here, $\mbox{E}$ is the total depth-averaged energy, APE is the available potential energy, $F$ and $D$ are the forcing and dissipation terms (from bottom drag, spectral filter at the small-scales, and/or from the implemented CNNs), and the fluxes are the transfers in spectral space. These tendency terms on the right hand side of the equation are diagnosed from the model runs, and a summary of these from the different model calculations over the same five year averaging period are shown in Fig.~\ref{fig:results_eddy_energy}. Starting first with the high resolution model truth (Fig.~\ref{fig:results_eddy_energy}$a$), positive values denote energy input at the relevant spatial scales. The generation of APE is from the imposed vertical shear and is always leading to an energy input, while bottom drag acts to remove energy across all scales. The flux of APE is from the large to small-scale, while the KE flux is upscale, consistent with theories of two-dimensional quasi-geostrophic turbulence \cite<e.g.>{Salmon80, JansenHeld14}. For comparison, while the low resolution model with no CNNs active display a qualitatively similar balance (Fig.~\ref{fig:results_eddy_energy}$b$; similar APE generation, but generally weaker fluxes and dissipation by bottom drag), there is a non-negligible balance at the small-scales between the KE flux related to accumulation at the grid-scale and the dissipation from the spectral filter, at odds with that displayed for the high resolution model truth.

With those observations in mind, it can be seen from Fig.~\ref{fig:results_eddy_energy}$c$,$d$,$e$ that the presence the trained CNNs, regardless of whether these are trained by offline or the online approach, lead to increased energy fluxes that are more in line with the high resolution model truth. All the resulting CNNs seem to be mimicking some sort of backscatter \cite<e.g.>{JansenHeld14}, as indicated by the brown lines being positive in the intermediate ranges away from the grid scale in Fig.~\ref{fig:results_eddy_energy}$c$,$d$,$e$; diagnoses indicate a backscatter in both KE and APE in comparable parts. The resulting CNNs also seem to be mimicking effects of the small-scale dissipation from the spectral filter, in that the explicit small-scale dissipation from the spectral filter (purple) line has reduced in magnitude relative to the no CNN case. Upon computing the integral over spectral space, the decrease in explicit dissipation arising from the spectral filter (Eq.~\ref{eq:sharp_filter}) in the ensemble mean relative to the no CNN low resolution model was diagnosed to be 42.7\% for the offline model, 64.4\% for the full online model, and 55.3\% for the approximately online model.

%--------------------------------------------------------
\subsection{Long-term similarity scores}

For a more qualitative measure of the skill in the models in terms of spectral similarity, we compute some of the metrics detailed in \citeA{Ross-et-al23}. For the distributions of the model variables (potential vorticity $q$, both components of velocity $u$ and $v$, streamfunction $\psi$, so four variables over two layers), we compute the 1-Wasserstein distance $W_1$ \cite<e.g.,>{Villani-OT}, which in the present case can be computed from the unsigned area difference between two cumulative distribution functions, i.e.,
\begin{linenomath*}
\begin{equation}\label{eq:dist diff}
  \mbox{distrib\_diff}(\mbox{sim1}, \mbox{sim2}; f) = \int^\infty_{-\infty}\left| P_{\rm sim1}(f\leq x) - P_{\rm sim2}(f\leq x) \right|\; \mathrm{d}x,
\end{equation}
\end{linenomath*}
where $f$ is the quantity of interest, and $P(f\leq x)$ is the empirically determined cumulative distributions, obtained from binning the variable data into a histogram based on their values over real space. For differences in distributions in spectral space (KE and enstrophy in both layers, KE flux, APE flux, APE generation and friction), we compute the root-mean-squared or $L^2$ distance of the spectral distributions, i.e.,
\begin{linenomath*}
\begin{equation}\label{eq:spectral diff}
  \mbox{spectral\_diff}(\mbox{sim1}, \mbox{sim2}; f) = \sqrt{\frac{1}{\mathcal{|K|}} \sum_{k \in \mathcal{K} } (f_{\rm sim1}(k)-f_{\rm sim2}(k))^2},
\end{equation}
\end{linenomath*}
where $f$ is the spectral distribution of the quantity of interest and is real, $k$ is the wavenumber, and the normalization is by $|\mathcal{K}|$, which is an isotropic wavenumber common to both simulations, up to 2/3 of the Nyquist frequency of the low resolution simulation. 

Once the distribution and spectral differences are computed, the similarity score is computed as
\begin{linenomath*}
\begin{equation}\label{eq:similarity}
  \mbox{similarity(hybrid, HR, diff}) = 1- \frac{\hbox{diff(hybrid,HR)}}{\hbox{diff(LR,HR)}},
\end{equation}
\end{linenomath*}
where HR, LR and hybrid denote diagnostics from the high resolution, low resolution with no CNN, and the corresponding hybrid models. The present relative measure is bounded above by 1, and the similarity score is maximized when the hybrid model matches the high resolution model truth exactly in the relevant metric. On the other hand, the ratio can be zero or even negative, achieved respectively when the hybrid model does no better or does worse than the low resolution model without the CNNs active.

\begin{figure}[tbhp]
\begin{center}
	\includegraphics[width=\textwidth]{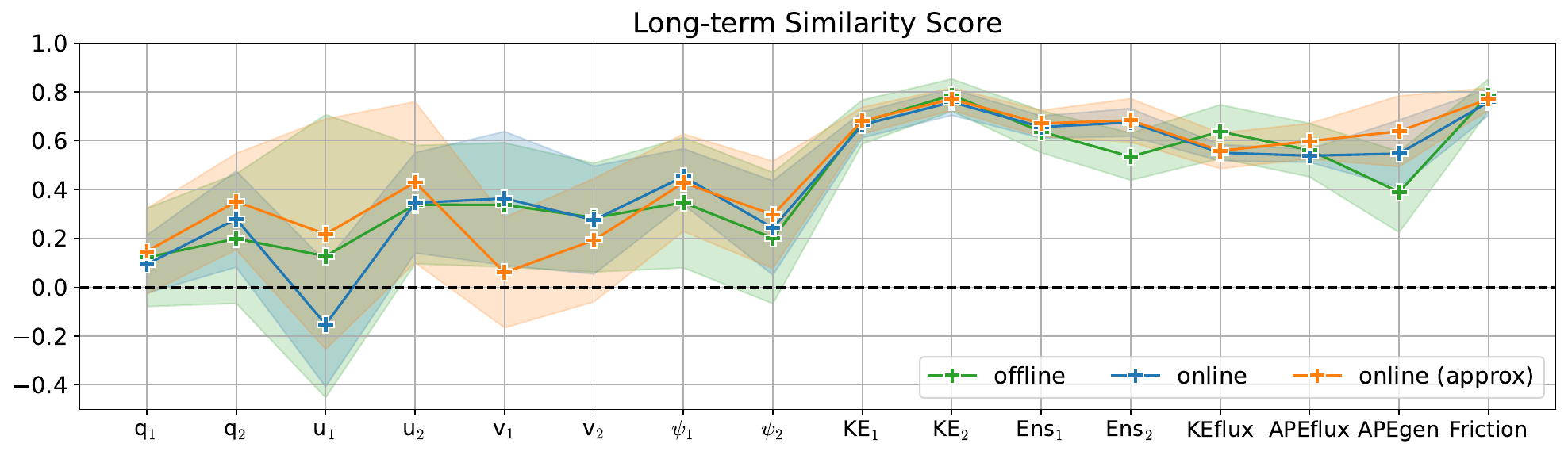}
	\caption{Distribution similarity scores for potential vorticity $q$, both components of the velocity $u$ and $v$, the streamfunction $\psi$, KE and enstrophy, depth-averaged KE flux, APE flux, APE generation, and the bottom drag (labeled as Friction). Subscripts 1 and 2 denote the upper and lower layer, for the (green) offline model, (blue) full online model, and (orange) approximately online model. The solid line is for the ensemble average over the ten members, and the shading denotes the standard deviation of the ensemble.}
	\label{fig:results_score_k}
\end{center}
\end{figure}

In Fig.~\ref{fig:results_score_k} we show the similarity scores for quantities of interest relating to the model. The model variable distributions have wide spreads, with biases particularly in the top layer (except for the streamfunction). The spectral distributions on the other hand show substantially less variability, and the inclusion of the CNNs generally lead to improvements, as indicated by a moderately positive similarity score. It is generally the case that the online models further improve on the offline models.

It is perhaps interesting to note that the lukewarm performance of the models in the variable distributions do not seem to have a strong imprint on the spectral fluxes, suggesting that `skill' is metric dependent, and one should probably not rely on a single metric for gauging model performance. The mismatch in the performance between model and spectral distributions may be due to the choice of loss function, which targets the sub-grid small-scale forcing terms $S^q$, so it may not be a surprise that we are obtaining improvements largely in the quadratic quantities that are strongly affected by the form of $S^q$. The choice of loss function was explored briefly, with some results reported in Sec.~\ref{choix-loss}. 

A final comment before further explorations with the online learning methodology, in the absence of the imposed hard constraint of net zero tendency of potential vorticity given by Eq.~\eqref{eq:constraint}, all previously reported results are still robust when the CNNs converge at the training stage, except for the similarity score (not shown). The resulting CNNs appear to lead to an accumulation of $q$ in the top layer for offline and approximately online model, leading to model blow up, as well as degradations in the similarity score (particularly in the model variable distributions). This phenomenon however does not seem to be present for the full online model case, which appears to be stable: even without the imposed hard constraint given in Eq.~\eqref{eq:constraint}, the accumulation of $q$ is non-zero but noticeably smaller than the resulting predictions from the offline and approximately online approach. This observation presumably points to the inherently stability properties from the use of a full adjoint-based online training strategy, and is something that will be further discussed in Sec.~\ref{conclusion}.

%-----------------------------------------------------------------------------

\section{Further comparisons of the online procedures}\label{online}

In this section we provide further explorations of the online learning methodology, motivated by some of the observations highlighted in the previous section, namely the dependence on the learning window size, the learning of small-scale dissipation, and possible flexibility in the choice of the loss function.

%%%

\subsection{Dependence on window size}\label{window_size}

One free parameter for the online learning procedure is the time window over which the training is performed. While we would suspect that the longer the window the better (because the model is then exposed to more of the model trajectory and data content), the choice also needs to be balanced by the computational cost and possible stability issues \cite<e.g., that reported in>{List-et-al22}.

\begin{figure}[tb]
\begin{center}
	\includegraphics[width=0.9\textwidth]{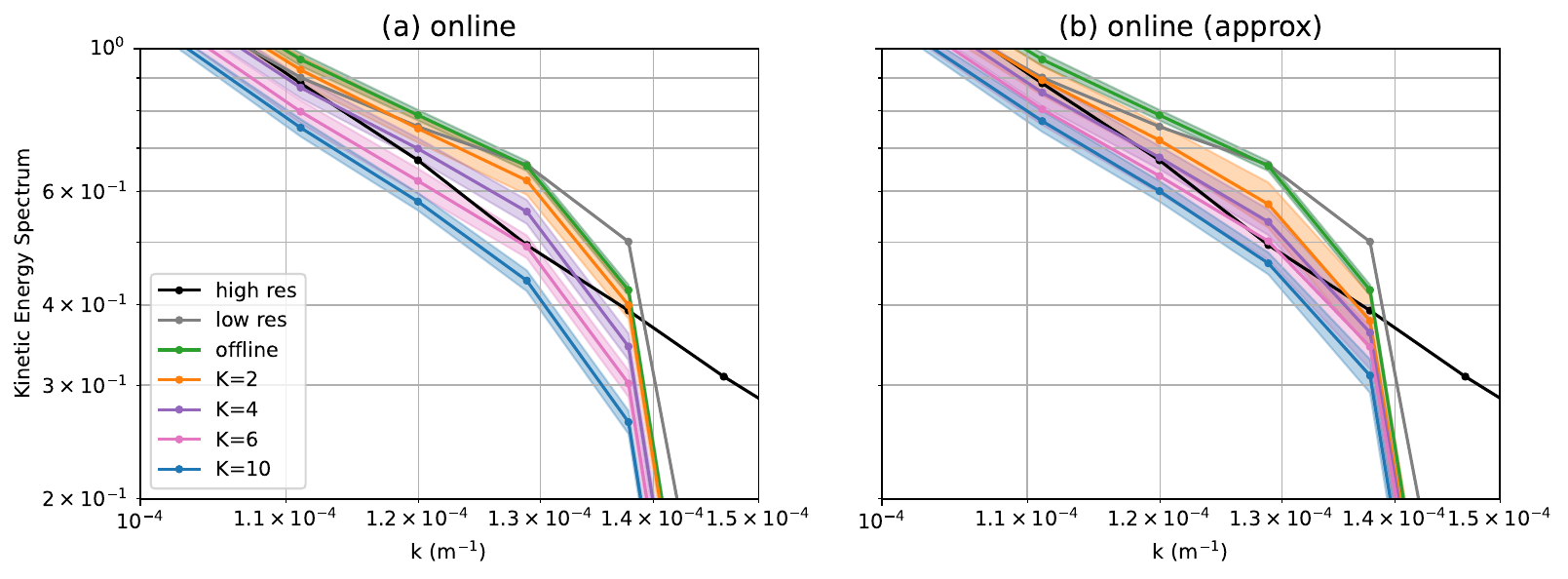}
	\caption{Zoomed in depth-averaged KE spectra (cf. Fig.~\ref{fig:results_spectral_ke}) focusing on the small scales for the ($a$) online model and ($b$) the approximately online model, for models trained with different time window sizes $K$. The solid line is for the ensemble average over the ten members, and the shading denotes the standard deviation of the ensemble.} 
	\label{fig:results_window}
\end{center}
\end{figure}

We have considered choices of window size on the performance of the resulting online model performance, and the main notable result we have found relates to the depth-averaged KE spectra at the small-scales, shown in Fig.~\ref{fig:results_window}. As a reminder, the models here are trained via a curriculum \cite<e.g.,>{List-et-al22, Frezat-et-al22, Ouala-et-al23, Kochkov-et-al24}, in that a mapping is first learnt from a window with restricted size, and that mapping is used as the initial mapping for learning at the next enlarged window size (cf. continuation approach in optimization and numerical methods). In both online approaches, there is the kick-back feature at the grid-scales when window size is small, but this feature gradually diminishes as window size increases. 

%The full online approach appears to have reached some sort of convergence (panel $a$). It is interesting to see that the approximately online approach appears to overshoot, resulting in a model having too little KE at the small-scales for the same default window size of ten time-steps; this may be one reason for a less than optimum skill of the associated model (orange line in Fig.~\ref{fig:results_score_k}). That the optimality of model skill for the approximately online approach occurs at the different location relative to the full online approach is consistent with the observations reported in \citeA{List-et-al24}, related to the truncation of parts of the back-propagation chain, modifying the resulting gradient calculations.

In the cases tested for the present set up, marginally longer window lengths beyond the default choice of ten time-steps lead to increases in training time, with not overly significant gains in model skill (not shown; largely similar to Fig.~\ref{fig:results_score_k}). An observation we make is that there appears to an overshoot with increased window size in Fig.~\ref{fig:results_window}. Note that Fig.~\ref{fig:results_window} is shown with a linear rather than logarithmic scale, so the overshoot is mild. The associated damping/backscatter near the grid-scale is not necessarily a negative feature, since it may be beneficial for model stability. We have not had noticeable issues with the gradient calculations associated with the back-propagation calculations, but this may be because the chosen training window size is rather small, and we have numerical stabilization from the presence of the imposed hard constraint in Eq.~\eqref{eq:constraint}, particularly for the approximately online case.

%%%

\subsection{CNN represented small-scale dissipation}\label{ssd}

\begin{figure}[tb]
\begin{center}
	\includegraphics[width=0.9\textwidth]{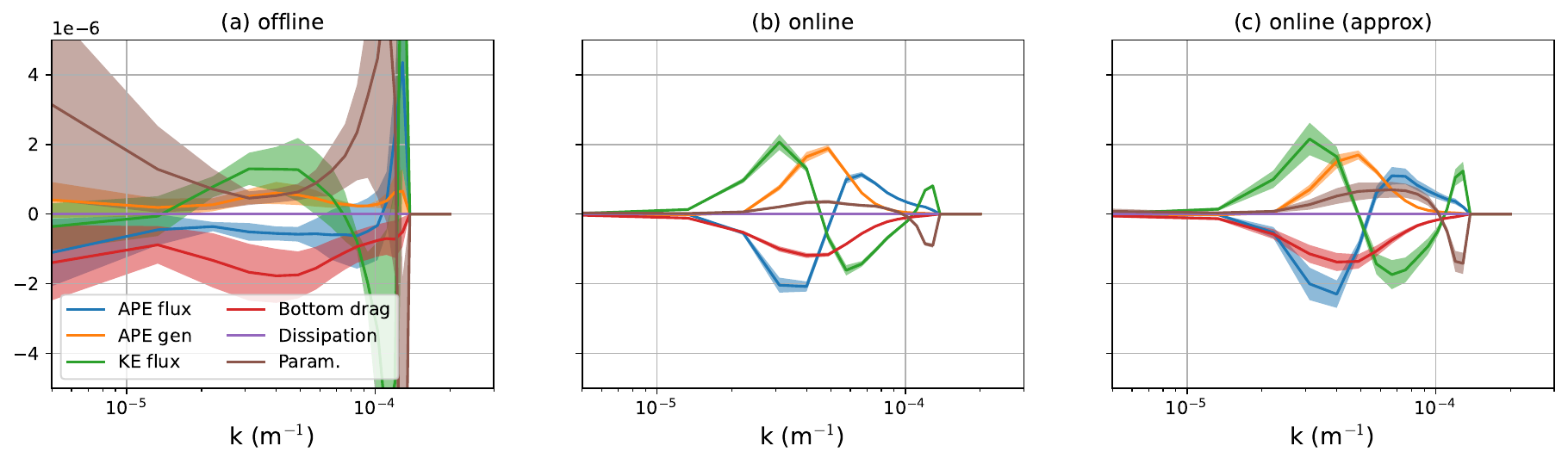}
	\caption{Spectral fluxes for \emph{a posteriori} calculations in a hybrid model with the explicit small-scale dissipation switched off, using the CNNs reported in Sec.~\ref{results_eddy}. ($a$) The offline hybrid model. ($b$) The online hybrid model. ($c$) The approximately online hybrid model. The colored lines are from the ten member ensemble averaged performance of the hybrid models, where shading denotes the standard deviation of the ensemble.}
	\label{fig:results_no_diss}
\end{center}
\end{figure}

The observation in the spectral energy budgets in Fig.~\ref{fig:results_spectral_ke} suggests that all the CNNs seem to have some representation of the backscatter and small-scale dissipation from the high resolution model. Here, we perform \emph{a posteriori} testing of the same hybrid models reported in Sec.~\ref{results_eddy} but in the absence of explicit small-scale dissipation, to see how the CNNs perform out-of-sample in this sense. The resulting spectral fluxes are shown in Fig.~\ref{fig:results_no_diss}. We see in Fig.~\ref{fig:results_no_diss}$a$ that the offline model performance is generally poor, with large biases over the whole bandwidth, and the ensemble spread is substantial. The spectral fluxes from the full online model shown in Fig.~\ref{fig:results_no_diss}$b$ is comparable in shape to Fig.~\ref{fig:results_eddy_energy}$d$ but with reduced magnitude, indicating a weaker energy cycle, but with some evidence of model skill when model is taken out-of-sample. The approximately online model results shown in Fig.~\ref{fig:results_no_diss}$c$ is also promising, although there is larger ensemble spread than the full online case. The similarity scores for the full and approximately online models are slightly lower (omitted for brevity). The observations here are robust for the full online model even in the absence of the hard constraint Eq.~\eqref{eq:constraint}; the approximately online model on the other hand has large biases (not shown), attributed to the accumulation of potential vorticity in the top layer.

For completeness, we also considered whether it was possible to turn off the explicit small-scale dissipation for online hybrid models at the \emph{training} stage, so that the online approaches would end up learning the grid-scale dissipation simply via exposure to the high resolution model data that resolves what would the grid-scales of the low resolution hybrid model. We note that the Orszag 2/3 dealising is still active, which imparts some numerical stability to the integration, but by itself is insufficient, in that the low resolution model without a CNN will crash in the same experimental setting due to grid-scale accumulation of energy and enstrophy. Both the full and approximately online methods will converge and return CNNs in this no dissipation setting, but the resulting model performance is generally poor, with a significant bias in the KE spectrum over a noticeable bandwidth in the small-scales (not shown). Reported observations are however for training based on a loss function defined with $S^q$; a curious result for an analogous procedure for a different loss function is given in the next subsection.

%%%

\subsection{Choice of loss function} \label{choix-loss}

In the results presented thus far the choice of the loss function targets the sub-grid forcing $S^q$ given in Eq.~\eqref{eq:sq0}, and for the present two-layer quasi-geostrophic system the CNNs are trained with separate loss functions defined per fluid layer. One choice that we could have made was to consider a combined loss function as in Eq.~\eqref{eq:lf_1loss}, choosing a coupling parameter $\alpha$ accordingly (or we could have considered appropriately normalized $\mathcal{L}_{1,2}$). Analogous training on the combined loss function does not lead to noticeable differences in the qualitative conclusions drawn so far, with similarity scores largely like that of Fig.~\ref{fig:results_score_k}, possibly with minor degradation in the skill. The minor degradation may simply be that the complexity (i.e. the number of degrees of freedom) of the CNNs trained on the one combined loss function is lower (since we kept the CNN architecture the same); this is however not something that we have investigated in comprehensive detail.

In the optimization literature, depending on the objective, the loss (or objective) function can often be designed or constructed to target different features. For example, instead of the usual root-mean-squared loss, the work of \citeA{GuillauminZanna21} considers the log-likelihood for their machine learning-based stochastic parameterization. The work of \citeA{Yan-et-al23} alludes to the use of negative Sobolev semi-norms as possible choices of loss functions to target larger-scale features \cite<cf. the works in the mixing literature such as>{Lin-et-al11, Thiffeault12}. For the present work, we note that the models display skill in the quantities relating to spectral fluxes, but less so in the model quantity variables (Fig.~\ref{fig:results_score_k}). The observations may have arisen from the fact that $S^q$, which is a small-scale variable relating to the nonlinear transfer term, was used in the definition of the loss function. In some sense $S^q$ is the only choice we have for forming the loss function for offline learning approaches, since we have no natural mapping of $S^q$ to the model variables, as the dynamical model $M: S_{q, t_0}(q_{t_0}| \boldsymbol{\theta}) \mapsto q_{t_0 + \Delta t}$ is not involved explicitly. This limitation is not present for online learning approaches, and one could consider using the model variables in the definition of the loss function as a way to target improvements in the model variable predictions \cite<e.g.,>{Frezat-et-al22, Ouala-et-al23}. One may suspect that if we are able to represent the model state well, then the statistics (e.g.,the implied $S^q$) may also be improved at the same time. The work of \citeA{Frezat-et-al22} finds that this was the case in a one-layer quasi-geostrophic model with full online learning with a loss function based on potential vorticity $q$.

We have attempted to train online models with loss functions that are solely in terms of model state variables (potential vorticity $q$ or streamfunction $\psi$), or as a mix of sub-grid and model state variables, e.g.,
\begin{linenomath*}
\begin{equation}\label{eq:Llmod}
  \mathcal{L}_{1,2} = \sum^K_{i=0} |S^q_{1,2;\ t_0+i\Delta t} - \hat{S}^q_{1,2;\ t_0+i\Delta t}|^2 + \tilde{\alpha} \sum^K_{i=0} |\overline{q}_{1,2;\ t_0+i\Delta t} - \hat{q}_{1,2;\ t_0+i\Delta t}|^2,
\end{equation}
\end{linenomath*}
for appropriate values of a weight parameter $\tilde{\alpha}$ (where we would expect the $\tilde{\alpha} \searrow 0$ limit to correspond to the primary scenario considered in the present work, although it is also possible that the problem here is singular). For the present case, and without comprehensive investigation into modifications of the CNN structure and/or hyper-parameters, online training (with explicit small-scale dissipation active in the hybrid model) of CNNs based on loss functions defined with the potential vorticity variable $q$ will converge, but the model performance is generally poor, with noticeable degradations in the KE spectrum, spectral fluxes and similarity scores (not shown). On the other hand, it is possible to train models with loss functions involving the streamfunction $\psi$, and the resulting models trained from loss functions defined in terms of $\psi$ have comparable similarity scores to that reported here in Fig.~\ref{fig:results_score_k}, but with minor improvements to the model distribution similarity scores. One possible reason could be that the streamfunction $\psi$, being a larger-scale field, is more amenable for extracting information from \cite<cf.>{Yan-et-al23}. However, the predicted $\hat{S}^q$ is not so comparable statistically to the diagnosed $S^q$, and the resulting spectral fluxes show large grid-scale dissipation in both KE and enstrophy, raising the question of whether one necessarily gets improvement in the spectral flux statistics even if the model state has reduced biases.

On the topic of stabilizers and constraints, one could build those into the mapping during the training stage, by modifying the CNN architecture or similar \cite<e.g.,>{Beucler-et-al21}, or modify the loss function accordingly. From the traditional optimization point of view, a constraint might be imposed by consider a loss function of the form $\mathcal{L} = \mathcal{L}_{\rm on} + \lambda\mathcal{C}$, where $\mathcal{C}$ is a constraint of interest. If we wanted to force the CNN here to output a prediction with zero mean, we might consider take $\mathcal{C}$ to be Eq.~\eqref{eq:constraint}, and force the constraint by the method of Lagrange multipliers, which involves augmenting the control variable from $\boldsymbol{\theta}$ to $(\boldsymbol{\theta}, \lambda)$, where $\lambda$ would be interpreted as the Lagrange multiplier. On the other hand, one could might consider imposing constraints approximately by viewing $\mathcal{C}$ instead as a penalization (e.g., $\mathcal{C} = \int (\hat{S}^q)^2\; \mathrm{d}A$ to keep the differentiability of the cost function); $\lambda$ would then be regarded as a penalization weight. The two approaches might be regarded as a hard and soft constraint optimization \cite<e.g.,>{Beucler-et-al21}, the benefit of the latter being that the control variable $\boldsymbol{\theta}$ is exactly as before with the same number of degrees of freedom. The same algorithms can be applied as normal but introducing subjectivity in the choice of $\lambda$, and the constraint is only enforced approximately. For the present case, instead of the hard constraint in Eq.~\eqref{eq:constraint}, we also considered imposing the zero-mean soft constraint as described above. The notable result is that we get similar results and stability properties as those reported thus far in this work, with the observation that the penalization approach appears to work better for the online than offline training (not shown). 

\begin{figure}[tbhp]
\begin{center}
	\includegraphics[width=0.9\textwidth]{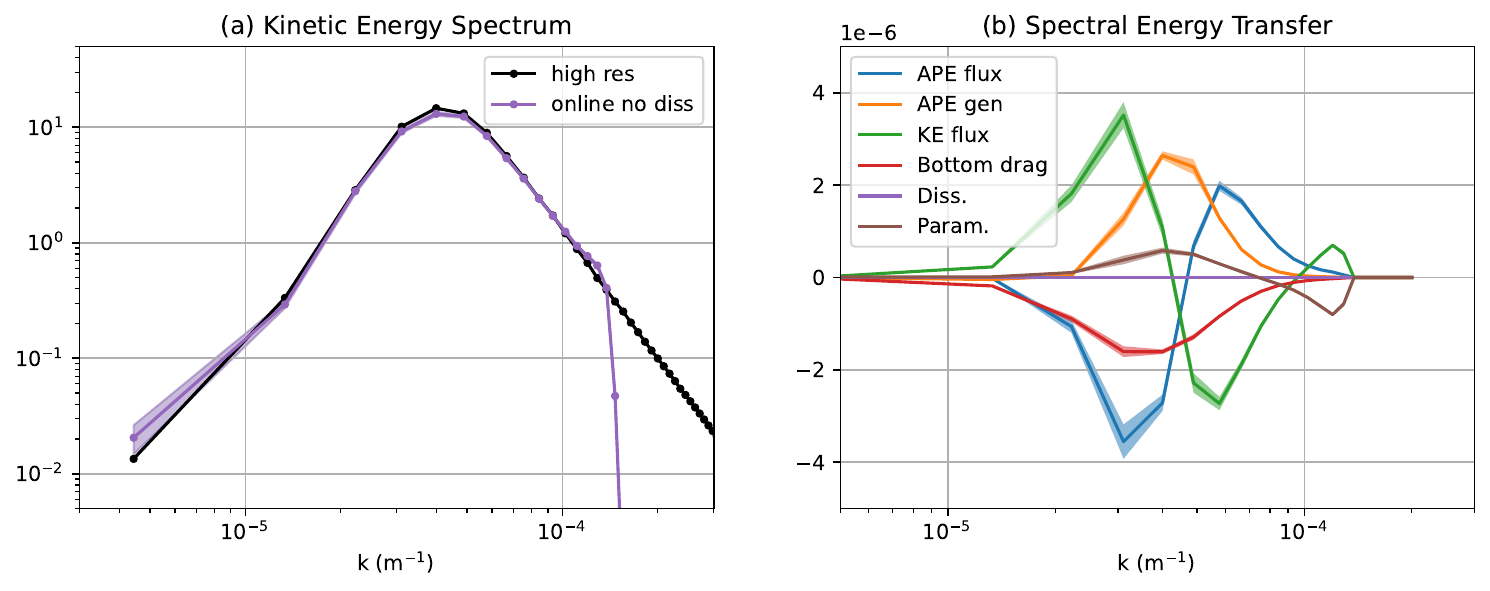}
	\caption{Diagnostics from the full online hybrid model, where the associated CNNs are trained with hybrid models \emph{without} any explicit small-scale dissipation active. ($a$) KE spectrum as Fig.~\ref{fig:results_spectral_ke}, and ($b$) spectral energy budget as Fig.~\ref{fig:results_eddy_energy}. The colored lines are from the ten member ensemble averaged performance of the hybrid models, where the shading denotes the standard deviation of the ensemble.}
	\label{fig:results_q_no_ssd}
\end{center}
\end{figure}

We close by noting that while it is true that model skill leaves a lot to be desired for online training with loss functions defined with the potential vorticity variable $q$ in hybrid models with explicit small-scale dissipation, some aspects are in fact improved for the full online learning procedure \emph{without} explicit small-scale dissipation in the hybrid models. Fig.~\ref{fig:results_q_no_ssd} shows the resulting KE spectra and spectral fluxes. While the resulting models show minor grid-scale kick-backs in Fig.~\ref{fig:results_q_no_ssd}$a$, a slightly reduced magnitude of energy cycles in Fig.~\ref{fig:results_q_no_ssd}$b$, and the resulting similarity scores are slightly worse than the offline model ones shown in Fig.~\ref{fig:results_score_k}, the model skill based on the considered metrics are otherwise not unreasonable (and even improving on the results shown in Fig.~\ref{fig:results_no_diss}). We do not have a convincing reason on why training without a numerical stabilizer leads to better model skill for loss functions defined with the potential vorticity variable $q$, and we leave this as an observation for now. The same procedure but for approximately online training fails at the training stage; this is not an unreasonable expectation, likely resulting from how the approximately online procedure of \citeA{List-et-al24} truncates the back-propagation calculations.

%-----------------------------------------------------------------------------

\section{Conclusion and discussion} \label{conclusion}

Building on the previous works demonstrating the efficacy of online learning in geophysical systems \cite<e.g.,>{Frezat-et-al22, Ouala-et-al23, Kochkov-et-al24}, known as temporal unrolling in other studies \cite<e.g.,>{List-et-al22, List-et-al24}, the present work provides an examination between online learning approaches and the more standard offline approach for eddy parameterization of baroclinic turbulence. The online approach involves the underlying fluid dynamical model during the training stage to provide a mapping from the predicted quantity (the sub-grid forcing $S^q$ in this work) to the model variable (the potential vorticity $q$ in this work) at the next time-step, and the control variables (the collection of neural network parameters, denoted $\boldsymbol{\theta}$ here) are adjusted so that the loss function defined over some time window is minimized. The online learning procedure considered in this work is derivative-based, and at its core is linked to methodologies in optimization and data assimilation, namely adjoint optimization \cite<e.g.,>{Gunzburger-control} and Four-Dimensional Variational Assimilation \cite<4DVAR; e.g.,>{Kalnay-DA} respectively. A plausible explanation for online learning is that the presence of the dynamical model forces the training to recognize that the target data provided has some temporal correlation/causality, exposing more information to the training procedure, and providing constraints of the solution space in which the trained model should reside. While the full adjoint-based online leading requires a differentiable dynamical model to enable the back-propagation calculations to obtain gradient information for the associated optimization problem, proposals exist to approximate the back-propagation and gradient calculations accordingly \cite<e.g.,>{Ouala-et-al23, List-et-al24}, or one could employ derivative-free methods \cite<e.g.,>{Iglesias-et-al13, KovachkiStuart19, LopezGomez-et-al22, Schneider-et-al22}; we considered here the approach put forward by \citeA{List-et-al24}.

For the present investigation, we extend on the previous work of \citeA{Frezat-et-al22} and \cite{Ouala-et-al23} for online learning in a one-layer quasi-geostrophic system to a two-layer quasi-geostrophic system that supports baroclinic dynamics, following the same testing framework of \citeA{Ross-et-al23}. We trained Convolution Neural Networks (CNN) with loss functions defined with the sub-grid forcing $S^q$ as defined in Eq.~\eqref{eq:sq0}, which is the residual between the filtered high resolution model truth and the low resolution hybrid dynamical model. The CNNs are trained via definitions of appropriate loss functions defined per fluid layer, and differ only in how they are trained, but with a hard constraint that the predicted $S^q$ should have zero mean given in Eq.~\eqref{eq:constraint}. Models are primarily tested in the \emph{a posteriori} sense, where the model skill is evaluated on their long term statistics in prognostic mode, which is a harder but more relevant test relating to parameterization \cite<e.g.,>{Srinivasan-et-al24}. For statistical significance, the testing was performed over a ten member ensemble of CNNs per training strategy. We do not impose controls on the back-propagation calculations since we have encountered issues like those reported in \citeA{List-et-al22}; this may be a combination of the problem set up, the hard constraint we imposed in Eq.~\eqref{eq:constraint}, and/or the small training window size that we took to be ten model time-steps.

We find that models trained with an online procedure lead to multiple benefits over those of offline models. The full online model skill does not depend sensitively on the presence of the hard constraint in Eq.~\eqref{eq:constraint}, while the approximately online model and the offline model do (not shown, but see descriptions in body text). The predicted sub-grid forcing for the full online model are closer in distribution to the high resolution model truth (see Fig.~\ref{fig:results_qqplot}). Both the full and approximately online models improve on the overall kinetic energy levels (Fig.~\ref{fig:results_spectral_ke}) and lead to more vigorous energy fluxes (Fig.~\ref{fig:results_eddy_energy} and \ref{fig:results_score_k}) in prognostic calculations. All the investigated CNNs seem to be learning some sort of backscatter that energizes the system, and some semblance of the grid-scale dissipation (Fig.~\ref{fig:results_eddy_energy}) regardless of learning approach. However, the offline models leads to persistent biases in the kinetic energy spectrum at the grid-scale, while the online models are able to remove that bias (Fig.~\ref{fig:results_spectral_ke}). There is additional evidence to suggest that the full online approach can lead to model skill when used out-of-sample, in tests where the hybrid models have no explicit small-scale dissipation (Fig.~\ref{fig:results_no_diss}). All of the aforementioned properties of the online learning approach are achieved for a rather small window length of ten time-steps (Fig.~\ref{fig:results_window}), indicating that we are able to obtain convergent training and obtain notable improvements in the model skill, for a modest investment in computational resources.

We have briefly explored whether including the the model variables explicitly into the definition of the loss functions to see whether we obtain improvements in the model state variables themselves for online learning \cite<cf. Fig.~\ref{fig:results_score_k} here; e.g.,>{Frezat-et-al22, Ouala-et-al23}. We find that models trained on the loss functions defined with the potential vorticity $q$, while stable, tends to have notable issues with skill (except for a curious case reported in Fig.~\ref{fig:results_q_no_ssd}). Models trained with loss functions defined by the streamfunction $\psi$ (related to $q$ via definition in Sec.~\ref{model}) have reasonable skill in the similarity scores. The issues with training convergence and/or model skill may be related to the instabilities associated with the back-propagation calculations, and could possibly be made more robust with stabilizations of the back-propagation calculations in the manner suggested by \citeA{List-et-al22}, and further imposition of constraints or conservation laws via appropriate design of loss functions or CNN architecture \cite<e.g.,>{Ling-et-al16b, Han-et-al19, Greydanus-et-al19, Beucler-et-al21}. The approximately online approaches could also be further tuned and/or refined \cite<e.g., taking more accurate/stable approximations for the gradient calculations; cf.>{Chen-et-al18, ChattopadhyayHassanzadeh23, Ouala-et-al23}. Loss functions beyond the usual root-mean-squared error is another possibility \cite<e.g., negative Sobolev semi-norms,>{Lin-et-al11, Thiffeault12, Yan-et-al23}. Further investigation is required to draw conclusive evidence on the benefit in utilizing the model variables to define loss function for online learning in the present baroclinic system.

For the present work, we presented results on a baroclinic turbulence model that was set up to be eddy dominated (cf. Fig.~\ref{fig:results_eddy_q}), via the choice of the bottom drag time-scale $r_{\rm ek}$ (see Table~\ref{tbn:pyqg}). It is known that strong zonal jets can form in rotating turbulence \cite<cf.>{Ross-et-al23, Srinivasan-et-al24}. Analogous investigations in the jet setting (by choosing a smaller value of $r_{\rm ek}$; not shown) seem to suggest that online learning still lead to some benefits, either via learning from scratch in the jet regime, or utilizing the CNNs trained in the eddy regime reported here but applied to hybrid model \emph{a posteriori} testing in the jet regime (i.e., applying the trained CNNs out-of-sample), both leading to moderately positive similarity scores. There are however additional subtleties with the training procedure such as the size of time window, and a strong dependence on the numerical model resolution (where the hybrid models' resolution at $64^2$ appears to fundamentally limit the model performance, but is noticeably alleviated when employing models at $128^2$ or higher). One key difference between the jet case and the eddy case is that there is the presence of a significant large-scale structure, and the hybrid models seem to have difficulty reducing the biases at the large-scale (as seen already in Fig.~\ref{fig:results_spectral_ke} of this work). There is likely room for improvement with further training with procedures indicated in the previous paragraph, but the qualitative findings are otherwise not overly different to the eddy case presented here, and the results in the jet case have been omitted for brevity.

To close, the present work primarily aims to compare at a rather coarse level on the benefits and drawbacks of the online versus offline learning approach, where online learning is found to show better skill in a range of metrics, allow for a more flexible in the definition of the loss function, results in CNNs that demonstrate a degree of skill when utilized out-of-sample, and likely results in CNNs that lead to more numerically stable integrations in prognostic calculations over a wider range of conditions. While the related investment in computational resources is likely worthwhile, in some ways the work has opened up more questions than it has answered. Further work into design and tuning choices of the machine learning models, stabilization techniques, choice and design of loss functions, enforcement of physical properties, and further investigations into the approximate online learning procedure (which does not require a differentiable model) or derivative-free methodologies should be considered, but is beyond the scope of the present work.

%-------------------------------------------------------------------------------

\section*{Data Availability Statement}

This work utilizes \texttt{pyqg-JAX} \cite<ver 0.8.1,>{Otness-et-al23, Otness24} and \texttt{Equinox} \cite{KidgerGarcia21}. The collection of code, model data and scripts used for generating the plots in this article are available
through the repository at \citeA{YanMak24}.

\acknowledgments

This study is a contribution to AI4PEX (Artificial Intelligence and Machine Learning for Enhanced Representation of Processes and Extremes in Earth System Models), a project supported by the European Union's Horizon Europe research and innovation programme under grant agreement no. 101137682, and also funded by the RGC General Research Fund 16304021. FEY and JM thanks James Maddison for technical comments for an earlier version of the work and article draft.

\end{document}